\documentclass[cameraready]{Interspeech}

\title{Trajectory Variance:\\ An Unsupervised Measure of Developmental Vocal Plasticity in Birdsong}

\author[affiliation={1}, orcid=0009-0001-0691-6664]{Kanghwi}{Lee}

\address{
    $^1$ Institute of Neuroinformatics, University of Zurich and ETH Zurich, Switzerland
}

\email{kanlee@ini.ethz.ch}


\usepackage{graphicx}

\begin{document}
\maketitle

\begin{abstract}
How much does a vocalization change over the course of development? We propose trajectory variance, a per-vocalization plasticity score that answers this question without type labels. A displacement model learns to predict age-conditioned shifts in autoencoder latent space; the variance of its predictions across target ages quantifies how much each vocalization would change if produced at different developmental stages. Evaluated on three zebra finches (183K--274K vocalizations, 40--101 days post-hatch), trajectory variance separates learned song syllables from innate calls (Cohen's d = 0.29--0.57, AUC = 0.58--0.67, after controlling for duration), while no nonparametric baseline achieves consistent separation. Trajectory variance also correlates with spectral flatness across all three birds (r = -0.48 to -0.75): more plastic vocalizations tend to have more tonal, structured spectra.
\end{abstract}

\section{Introduction}

How much does a vocalization change over the course of development? Static acoustic descriptors---spectral shape, duration, energy---characterize what a sound is at the moment it is produced, but not how it evolves. Analogous questions arise in single-cell genomics, where optimal-transport methods track how cell states shift along developmental trajectories~\cite{Schiebinger2019}. We propose \textbf{trajectory variance}, a measure that applies counterfactual reasoning to vocal development: if this vocalization had been produced at a different age, how different would it be?

The approach learns a model that predicts age-conditioned shifts in a latent representation of vocalizations. For each vocalization, counterfactual predictions are generated at several target ages, and the variance of these predictions serves as a plasticity score. Vocalizations predicted to change substantially across ages receive high trajectory variance; those predicted to remain stable receive low variance. No vocalization-type labels or manual annotation are needed.

We validate this approach on zebra finch vocal development---a well-studied model system where learned song syllables coexist with innate calls \cite{Brainard2002, Zann1996}. During a critical period from roughly 25 to 90 days post-hatch (dph), juvenile birds progress from unstructured subsong through plastic song to a crystallized adult motif \cite{Tchernichovski2001}. Individual syllables are shaped by local exploration \cite{Ravbar2012} and gradually stabilized \cite{Vallentin2016}, while innate calls remain comparatively unchanged. Existing computational approaches classify vocalizations by their static acoustic form \cite{Tchernichovski2000, Cohen2022, Koch2025, Sainburg2020, Goffinet2021}, or characterize developmental dynamics through repertoire dating \cite{Kollmorgen2020} and generative modeling \cite{Brudner2023}. Trajectory variance complements these: rather than classifying what a sound is or when similar sounds were produced, it measures how much a vocalization is predicted to change.

Applied to three birds (183K--274K vocalizations each, 40--101~dph), trajectory variance from our displacement model reliably separates song syllables from calls with small-to-medium effect sizes (Cohen's $d = 0.29$--$0.57$, AUC $= 0.58$--$0.67$ after controlling for duration), while no nonparametric baseline---Gaussian OT, per-age $k$-NN, or per-age optimal assignment---achieves consistent separation across all birds. Trajectory variance also correlates with spectral flatness across all three birds ($r = -0.48$ to $-0.75$).

Our contribution is not an efficient song/call classifier---simpler methods suffice---but a per-vocalization measure of how much a vocalization is predicted to change over development.

\section{Related Work}

\textbf{Birdsong analysis and development.} Quantitative birdsong analysis has progressed from hand-crafted features~\cite{Tchernichovski2000} through supervised segmentation~\cite{Cohen2022, Koch2025} to unsupervised latent-space methods~\cite{Goffinet2021, Sainburg2020}. These characterize vocalization structure at a given moment. Temporal dynamics of development have been studied through repertoire dating---tracking when a vocalization's nearest neighbors were produced~\cite{Kollmorgen2020}---and through generative models linking vocal variability to circadian and developmental factors~\cite{Brudner2023}. Our work is complementary: rather than classifying sounds or dating their neighbors, we ask how a vocalization would differ if produced at another developmental age.

\textbf{Optimal transport in generative modeling.} Flow matching~\cite{Lipman2023} and rectified flow~\cite{Liu2023} learn continuous normalizing flows via conditional vector fields, and minibatch OT coupling~\cite{Tong2024} improves training stability. Our model uses OT coupling to form training pairs but predicts displacements directly rather than integrating a velocity field, resulting in a single-pass inference without ODE solvers.

\textbf{Counterfactual reasoning.} The question ``what would this observation look like under different conditions'' is central to causal inference~\cite{Pearl2009} and applied in voice conversion and speech processing. To our knowledge, counterfactual age prediction has not been applied to animal vocal development.

\section{Method}

\begin{figure*}[t]
  \centering
  \includegraphics[width=\linewidth]{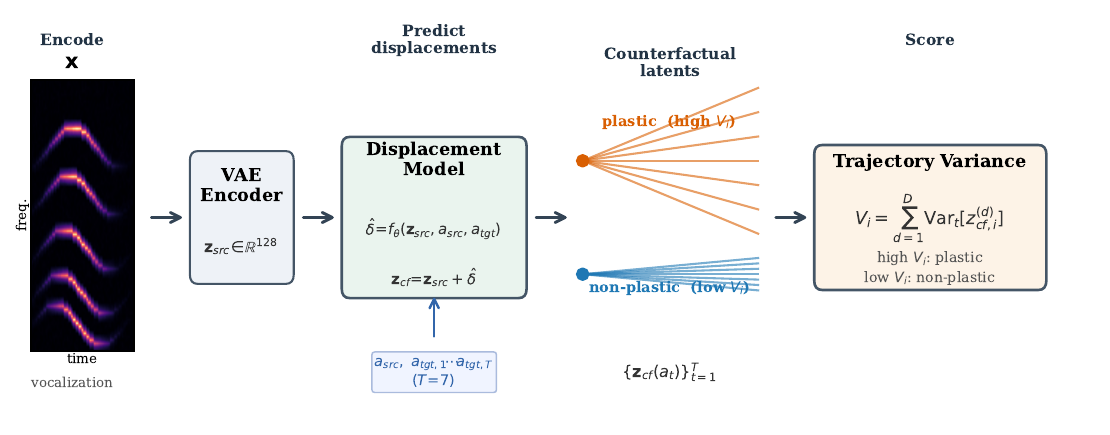}
  \caption{Pipeline overview. A VAE encodes each vocalization spectrogram into a 128-dimensional latent vector $\mathbf{z}_{src}$. The displacement model $f_\theta$, conditioned on source and target ages, predicts a latent shift $\hat{\boldsymbol{\delta}}$ for each of $T\!=\!7$ target ages. Trajectory variance $V_i$ is the summed per-dimension variance of the resulting counterfactual latents: learned vocalizations (developmentally plastic) produce wide fans; innate vocalizations produce narrow ones.}
  \label{fig:pipeline}
\end{figure*}

Our pipeline (Fig.~\ref{fig:pipeline}) consists of three stages: a variational autoencoder compresses spectrograms into a latent space, a displacement model learns age-conditioned latent displacements, and trajectory variance is computed from counterfactual predictions across target ages.

\subsection{Dataset}

Vocalizations were recorded from three juvenile male zebra finches during vocal development (40--101 dph) and segmented into individual vocalizations without type labels. Spectrograms were extracted from a pre-existing data pipeline (\SI{32}{\kilo\hertz} sampling rate, 512-sample window, 128-sample hop length, and 123 linear frequency bins after filtering low-frequency noise), right-padded to 100 time frames ($\approx$\,400\,ms), and z-score normalized per bird. Dataset sizes: 222K (Bird~A), 274K (Bird~B), 183K (Bird~C).\footnote{Code, models, and data: \url{https://github.com/hwiora/trajectory_variance}. Raw recordings are available from the author on request.}

\subsection{Spectrogram VAE}

A convolutional variational autoencoder \cite{KingmaWelling2014} compresses each spectrogram $\mathbf{x} \in \mathbb{R}^{123 \times 100}$ to a latent vector $\mathbf{z} \in \mathbb{R}^{128}$. The encoder applies three \texttt{Conv1d} layers ($123 \!\to\! 128 \!\to\! 256 \!\to\! 512$ channels, kernel size~3, stride~2) with batch normalization and GELU activations, followed by separate linear projections to $\boldsymbol{\mu}$ and $\log\boldsymbol{\sigma}^2 \in \mathbb{R}^{128}$. The decoder mirrors this architecture with transposed convolutions. The loss combines masked MSE reconstruction (ignoring padding), KL divergence (weight $10^{-3}$), and $L_2$ latent regularization (weight $10^{-4}$). We use AdamW \cite{Loshchilov2019} ($\text{lr}=3 \!\times\! 10^{-4}$), cosine annealing, 100 epochs, batch size 128. One VAE is trained per bird.

\subsection{Age-conditioned displacement model}

Given encoded latents $\{(\mathbf{z}_i, a_i)\}_{i=1}^{N}$, where $a_i$ denotes the developmental age (in days post-hatch) at which vocalization $i$ was recorded, we train a model to predict the displacement $\boldsymbol{\delta} = \mathbf{z}_{\mathrm{tgt}} - \mathbf{z}_{\mathrm{src}}$ for vocalizations resampled across ages. Because no longitudinal recordings track the same vocalization over time, training pairs are formed via mini-batch optimal transport: within each batch, we compute the pairwise squared Euclidean cost between source- and target-age latents and solve the assignment with the Hungarian algorithm \cite{Kuhn1955}.

The displacement model is a 6-layer residual MLP with adaptive layer normalization (AdaLN) \cite{Peebles2023}, conditioned on source and target ages via sinusoidal embeddings. The output layer is zero-initialized so that the displacement is near-zero at initialization. The model predicts displacements in a single forward pass:
\begin{equation}
  \mathbf{z}_{\mathrm{cf}} = \mathbf{z}_{\mathrm{src}} + f_\theta(\mathbf{z}_{\mathrm{src}},\, a_{\mathrm{src}},\, a_{\mathrm{tgt}}).
  \label{eq:displacement}
\end{equation}
Latents are z-score normalized per dimension; ages are normalized to $[\epsilon, 1\!-\!\epsilon]$ ($\epsilon = 0.05$). Training minimizes MSE on displacements. Optimization: AdamW ($\text{lr}\!=\!3 \!\times\! 10^{-4}$, weight decay $10^{-4}$), cosine annealing, gradient clipping at 1.0, 200 epochs, batch size 256, with model selection on 10\% held-out validation loss.

\subsection{Trajectory variance}

For each vocalization $\mathbf{z}_i$, we generate counterfactuals at $T\!=\!7$ evenly spaced target ages spanning the full developmental range. Trajectory variance is defined as:
\begin{equation}
  V_i = \sum_{d=1}^{D} \mathrm{Var}_{t=1}^{T}\bigl[ z_{\mathrm{cf},i}^{(d)}(a_t) \bigr],
  \label{eq:traj_var}
\end{equation}
where $z_{\mathrm{cf},i}^{(d)}(a_t)$ is the $d$-th latent dimension of the counterfactual at target age $a_t$. High $V_i$ indicates a vocalization is predicted to change substantially across development; low $V_i$ indicates stability.

\section{Experiments}

All analyses are applied independently to each bird. We first label each vocalization as song or call, then evaluate (1)~whether trajectory variance correlates with known acoustic markers of vocal development, (2)~whether it separates song syllables from calls, and (3)~whether the learned displacement model outperforms simpler baselines.

\subsection{Song/call labels}

We label vocalizations using a temporal bout heuristic rather than acoustic clustering, avoiding circularity with the acoustic features we evaluate against. Consecutive vocalizations separated by $<200$\,ms gaps are grouped into bouts; vocalizations in bouts of three or more are labeled \textit{song}, and all others \textit{call}. This exploits the rapid-fire sequencing that characterizes zebra finch song practice \cite{Tchernichovski2001}, without reference to spectral content. Manual inspection of randomly sampled bouts yielded approximately 83\% agreement with this heuristic.

\subsection{Correlations with acoustic features}

Table~\ref{tab:results} reports Pearson correlations between trajectory variance and two acoustic features: spectral flatness---mathematically equivalent to Wiener entropy \cite{Tchernichovski2000}---and vocalization duration. Duration correlates strongly across all three birds ($r = 0.70$--$0.80$). Because song syllables are generally longer than calls, this correlation means that raw trajectory variance could separate song syllables from calls simply by reflecting duration differences. To test whether the model captures developmental structure beyond length, the song/call separation metrics in Table~\ref{tab:baselines} ($d_r$, AUC) are computed on trajectory variance residualized on duration. Spectral flatness correlates with trajectory variance across all three birds ($r = -0.48$ to $-0.75$; Table~\ref{tab:results}); lower flatness---more tonal spectra---accompanies higher trajectory variance. We return to this relationship in the Discussion.

\begin{table}[t]
  \caption{Pearson correlation ($r$) between trajectory variance and acoustic features (3K evaluation subset per bird). All spectral-flatness and duration correlations are significant ($p < 10^{-10}$).}
  \label{tab:results}
  \centering
  \small
  \begin{tabular}{l c c c}
    \toprule
    \textbf{Feature} & \textbf{A} & \textbf{B} & \textbf{C} \\
    \midrule
    Spectral flatness  & $-.48$ & $-.75$ & $-.60$ \\
    Duration           & $+.70$ & $+.70$ & $+.80$ \\
    \bottomrule
  \end{tabular}
\end{table}

\begin{figure*}[t]
  \centering
  \includegraphics[width=\linewidth]{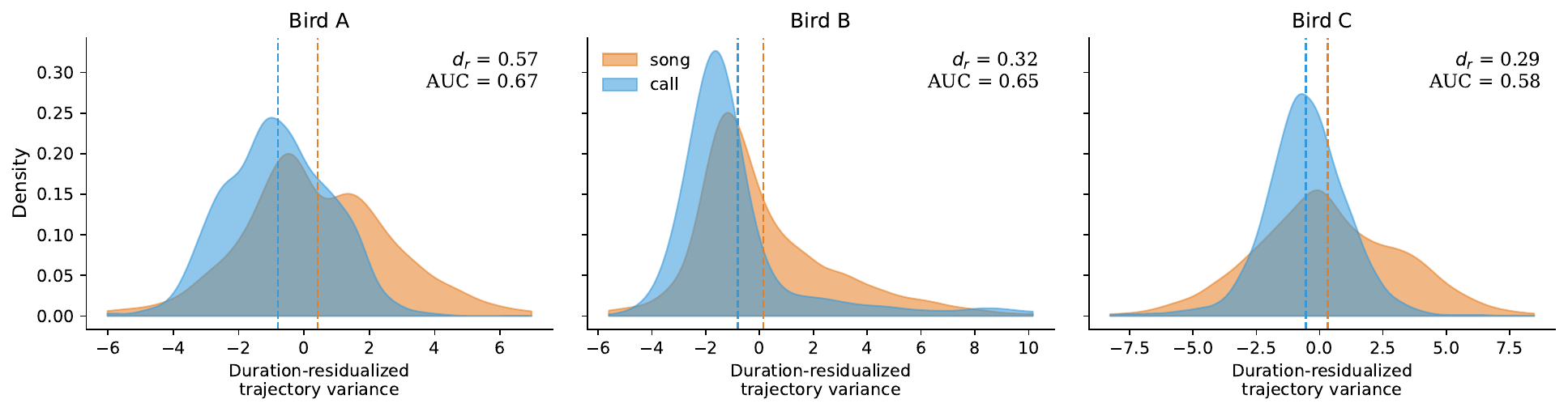}
  \caption{Distribution of duration-residualized trajectory variance for song (orange) and call (blue) vocalizations (10K samples per bird). Dashed lines mark group means. Song vocalizations receive consistently higher trajectory variance after controlling for duration, with effect sizes decreasing from Bird~A ($d_r\!=\!0.57$) to Bird~C ($d_r\!=\!0.29$).}
  \label{fig:songcall}
\end{figure*}

\subsection{Baseline comparison}

We compare the learned displacement model against three nonparametric baselines, all operating in the same z-score normalized latent space with the same $T\!=\!7$ target ages and the same evaluation subset (3K samples per bird):

\begin{itemize}
  \item \textbf{Gaussian OT}: Fits a Gaussian to the latent vectors at each age and applies the closed-form optimal-transport map~\cite{Peyre2019} between the source- and target-age distributions. The resulting affine transform is identical for all vocalizations at a given source age, capturing only population-level drift.
  \item \textbf{Per-age $k$-NN ($k\!=\!10$)}: Mean of $k$ nearest neighbors at each target age. Captures local structure but lacks developmental directionality.
  \item \textbf{Per-age OT}: Hungarian assignment between source- and target-age pools. One-to-one matching avoids the averaging of per-age $k$-NN but depends on pool composition.
\end{itemize}

Table~\ref{tab:baselines} reports two song/call separation metrics per method: $d_r$, Cohen's $d$ on duration-residualized trajectory variance; and AUC, the area under the ROC curve for the same comparison. Figure~\ref{fig:songcall} shows the corresponding distributions for the displacement model. The displacement model is the only method yielding consistently positive $d_r$ across all three birds ($d_r = 0.29$--$0.57$, AUC $= 0.58$--$0.67$), with small-to-medium effect sizes. No baseline achieves consistent separation: Gaussian OT is weakly positive for two birds but negative for Bird~B ($d_r = -0.16$); per-age $k$-NN produces near-zero $d_r$ across all birds (AUC $\approx 0.50$), indicating chance-level separation; and per-age OT shows a notable reversal ($d_r = -0.19$ to $-0.41$, AUC $= 0.39$--$0.45$), assigning higher trajectory variance to calls than to song syllables. This reversal likely reflects the sensitivity of one-to-one matching to pool composition: when age bins are unevenly populated, stereotyped calls may be matched to distant targets, inflating their apparent variance.

\begin{table}[t]
  \caption{Song/call separation: $d_r$~=~Cohen's $d$ on duration-residualized trajectory variance; AUC~=~area under ROC curve (3K subset per bird).}
  \label{tab:baselines}
  \centering
  \small
  \setlength{\tabcolsep}{4pt}
  \begin{tabular}{l@{\hspace{8pt}} c c c @{\hspace{8pt}} c c c}
    \toprule
    & \multicolumn{3}{c}{$d_r$} & \multicolumn{3}{c}{AUC} \\
    \cmidrule(lr){2-4} \cmidrule(lr){5-7}
    & A & B & C & A & B & C \\
    \midrule
    Displacement (ours) & \textbf{+.57} & \textbf{+.32} & \textbf{+.29} & \textbf{.67} & \textbf{.65} & \textbf{.58} \\
    Gaussian OT         & +.06 & $-.16$ & +.07 & .52 & .44 & .52 \\
    Per-age $k$-NN      & $-.05$ & $-.07$ & $-.03$ & .50 & .49 & .46 \\
    Per-age OT          & $-.19$ & $-.41$ & $-.24$ & .45 & .39 & .42 \\
    \bottomrule
  \end{tabular}
\end{table}

\section{Discussion}

\subsection{What trajectory variance captures}

Trajectory variance from the displacement model separates song syllables from calls across all three birds (Table~\ref{tab:baselines}, Fig.~\ref{fig:songcall}), without any acoustic feature engineering or vocalization labels. The effect sizes are small-to-medium ($d_r = 0.29$--$0.57$), consistent with the expectation that developmental plasticity is one of several factors distinguishing song syllables from calls, rather than a perfect separator. Across all three birds, trajectory variance correlates negatively with spectral flatness (Wiener entropy~\cite{Tchernichovski2000}; $r = -0.48$ to $-0.75$; Table~\ref{tab:results}): more plastic vocalizations have lower flatness---more tonal, structured spectra. Together, these results indicate that trajectory variance provides a single, label-free, per-vocalization measure of predicted developmental change---one that both recovers the song/call distinction and aligns with a classical spectral descriptor.

\subsection{Learned vs.\ nonparametric baselines}

The baselines span a range of approaches: Gaussian OT applies the same affine map to all vocalizations at a given age (population-level drift only); per-age $k$-NN retrieves the most similar observed vocalizations at the target age; per-age OT constructs one-to-one correspondences. The learned displacement model conditions on both the source latent and the age gap, producing nonlinear, sample-specific predictions. Its advantage in song/call separation (Table~\ref{tab:baselines}: $d_r > 0$ for all birds, vs.\ near-zero or negative for all baselines) confirms that this conditioning captures developmental structure that population-level or retrieval-based methods miss. The per-age OT reversal ($d_r < 0$; calls assigned higher variance than song syllables) is particularly informative: one-to-one matching depends heavily on pool composition and age-bin boundaries, and does not model how a specific vocalization would change---only which other vocalization it most resembles. Per-age $k$-NN, despite capturing local latent structure, produces chance-level song/call separation ($d_r \approx 0$, AUC $\approx 0.50$), indicating that retrieval of similar vocalizations at target ages does not yield meaningful plasticity estimates.

\subsection{Limitations}

\textbf{Duration correlation.} Duration correlates strongly with trajectory variance ($r = 0.70$--$0.80$; Table~\ref{tab:results}), likely because the VAE's fixed-length spectrograms (right-padded to 100 frames) implicitly encode vocalization length. Since song syllables are longer than calls in general, raw trajectory variance would trivially separate them. The song/call metrics ($d_r$, AUC) in Table~\ref{tab:baselines} are therefore residualized on duration, confirming that the displacement model captures structure beyond length.

\textbf{Relation to static descriptors.} Trajectory variance correlates with spectral flatness across all three birds ($r = -0.48$ to $-0.75$; Table~\ref{tab:results}), though these correlations are partial, leaving substantial variance unexplained. Whether the remaining variation reflects developmental information beyond static spectral form is not resolved here and is a direction for future work.

\textbf{No longitudinal ground truth.} Our evidence is cross-sectional. The transport model predicts what a vocalization would look like at other ages based on the population, not what it actually became in a given individual. We cannot directly verify these counterfactual predictions.

\textbf{Label heuristic.} The bout-based song/call labels rely on temporal proximity, not expert annotation. This avoids circularity with acoustic features but introduces its own biases: isolated song syllables are mislabeled as calls, and innate calls interspersed within song bouts as song.

\textbf{Generation quality.} As a sanity check, we compute Fr\'{e}chet Audio Distance (FAD)~\cite{Kilgour2019} between decoded counterfactual spectrograms and real recordings (Griffin-Lim inversion, VGGish embeddings). Pooled FAD ranges from 0.01 to 0.06 across birds. For reference, FAD between two independent halves of the real data is 0.002--0.007, so the counterfactuals are distinguishable from real recordings. We report FAD only as a decoder-fidelity diagnostic, not a perceptual claim; trajectory variance operates entirely in latent space and does not depend on the perceptual quality of decoded spectrograms.

\textbf{Generalizability.} We study three birds from one colony and species. Extension to other species or to human vocal development~\cite{Lipkind2013} remains untested.

\section{Conclusion}

We proposed counterfactual age displacement as a method for measuring developmental vocal plasticity without vocalization type labels. Applied to three zebra finches (183K--274K vocalizations, 40--101 dph), the learned displacement model reliably separates song syllables from calls with small-to-medium effect sizes (Cohen's $d = 0.29$--$0.57$, AUC $= 0.58$--$0.67$ after controlling for duration), while no nonparametric baseline achieves consistent separation. Trajectory variance also correlates with spectral flatness ($r = -0.48$ to $-0.75$), with plasticity associated with more tonal spectra. The approach is limited by the absence of longitudinal ground truth---and the raw measure is confounded by vocalization duration ($r = 0.70$--$0.80$), which we address by residualization---but provides a complementary perspective to static acoustic descriptions for studying vocal development.

\section{Acknowledgments}

\ifcameraready
  The author thanks Dina Lipkind for sharing the zebra finch recordings, Richard H.R.\ Hahnloser for his support, and colleagues in the Birdsong and Natural Language Group for helpful discussions. This research was funded by the Swiss National Science Foundation, Projects 31003A\_182638 and 205320\_215494/1.
\fi

\section{Generative AI use disclosure}

The original code, data analysis, and manuscript drafts were produced entirely by the author. Generative AI tools (Claude, GitHub Copilot) were used solely for editing and polishing the manuscript and formatting code. All scientific claims, experimental design, data interpretation, and final content are the sole responsibility of the author.

\bibliographystyle{IEEEtran}
\bibliography{mybib}

\end{document}